\documentclass[superscriptaddress,aps,twocolumn,showpacs,showkeys,notitlepage,amsmath, amssymb,floatfix]
{revtex4-1}

\usepackage{graphicx}
\usepackage[mathlines]{lineno}
\usepackage[svgnames]{xcolor}
\usepackage{hyperref}
\usepackage{color}
\hypersetup{
	colorlinks = true,
	linkcolor = Blue,
	citecolor = Blue,
	urlcolor  = Blue
}
\newcommand{\msr}{$\mu$SR}
\usepackage{comment}
\usepackage{ltablex}
\usepackage{threeparttable}
\usepackage[referable]{threeparttablex}

\begin{document}

\title{Quantum effects in muon spin spectroscopy within the stochastic self-consistent harmonic approximation}

\author{Ifeanyi John Onuorah}
\email[]{ifeanyijohn.onuorah@unipr.it}
\affiliation{Department of Mathematical, Physical and Computer Sciences, University of Parma, Italy}
\author{Pietro Bonf\`{a}}
\affiliation{Department of Mathematical, Physical and Computer Sciences, University of Parma, Italy}
\affiliation{Centro S3, {CNR-Istituto} Nanoscienze, 41125 Modena, Italy}
\author{Roberto De Renzi}
\email[]{roberto.derenzi@unipr.it}
\affiliation{Department of Mathematical, Physical and Computer Sciences, University of Parma, Italy}
\author{Lorenzo Monacelli}
\affiliation{Dipartimento di Fisica,Universit\`{a} di Roma Sapienza, Italy}
\author{Francesco Mauri}
\affiliation{Dipartimento di Fisica,Universit\`{a} di Roma Sapienza, Italy}
\author{ Matteo Calandra}
\affiliation{Sorbonne Universit\`{e}, CNRS, Institut des Nanosciences de Paris, UMR7588, F-75252 Paris, France}
\author{Ion Errea}
\affiliation{Fisika Aplikatua 1 Saila, Gipuazkoako Ingeniaritza Eskola, University of the Basque Country (UPV/EHU), Donostia-San Sebastian, Basque Country, Spain}
\affiliation{Centro de F\'isica de Materiales (CSIC-UPV/EHU),  Donostia-San Sebastian, Basque Country, Spain}
\affiliation{Donostia International Physics Center (DIPC), Donostia-San Sebastian, Basque Country, Spain}

\date{\today}

\begin{abstract}
Muon spin rotation experiments involve muons that experience zero-point vibration at their implantation sites. Quantum-mechanical calculations of the host material usually treat the muon as a point impurity, ignoring its zero-point vibrational energy that, however, plays a role in determining the stability of calculated implantation sites and estimating physical observables. As a first-order correction, the muon zero-point motion is usually described within the harmonic approximation, despite the anharmonicity of the crystal potential. Here we apply the stochastic self-consistent harmonic approximation, a quantum variational method devised to include anharmonic effects in total energy and vibrational frequency calculations, in order to overcome these limitations and provide an accurate ab initio description of the quantum nature of the muon. We applied this full quantum treatment to the calculation of the muon contact hyperfine field in textbook-case metallic systems, such as  Fe, Ni, Co including  MnSi and MnGe, improving agreement with experiments. Our results show that there are anharmonic contributions to the muon vibrational frequencies with the muon zero-point energies above 0.5\,eV.  Finally, in contrast to the harmonic approximation, we show that including quantum anharmonic fluctuations, the muon stabilizes at the octahedral site in bcc Fe.    
\end{abstract}


\maketitle

\section{\label{sec:anharmcalc}Introduction}

In muon spin rotation (\msr) experiments, spin-polarized positive (anti)muons are used to probe the microscopic field distribution at the interstitial site(s) where the $\mu^{+}$ stop inside the sample under investigation. The extreme sensitivity of the muon to small magnetic fields as well as the absence of quadrupolar coupling makes this technique very effective in probing magnetic orders, offering a valuable alternative to neutron scattering. This approach, which shares many similarities with nuclear magnetic resonance, has the advantage of being applicable to virtually any material, but it has the drawback that the interstitial sites where the muon stops and the nature of muon interaction with the host are generally unknown. Here we discuss an improved method to tackle this problem based on computational chemistry methods.

An accurate, \textit{ab initio}, description of the electron-muon interaction in periodic solids has been out of reach until a few years ago. The dramatic increase of both the computational power and the accuracy of first-principles calculations make this goal possible. Self-consistent electronic structure calculations, in particular those based on density functional theory (DFT), are already employed to study the muon implantation site, muon interaction parameters, and for understanding the muon-induced distortion in the lattice~\cite{rath1979,moller2013b,moller2013a,bonfa2013,bonfa2016,onuorah2018,liborio2018}. This turns out to be a very valuable tool for analyzing experimental data and interpreting the results~\cite{disseler2014}.  The knowledge of the muon implantation site(s) and of the hyperfine field allows very important quantitative information, including the magnetic structure and the moment size, to be obtained from \msr\ experiments. 
Moreover, a reliable quantum calculation of the muon embedded in the system under investigation provides an estimate for its induced perturbation; the probe is an impurity and it may in principle alter the local electronic properties. Fortunately this is a very rare case, and yet assessing these rare cases \cite{Foronda2015,dalmas2018} is very important.

However, self-consistent DFT calculations often treat the muon as just another atom in the lattice,  within the Born-Oppenheimer (BO) approximation~\cite{born1927}, without taking into consideration the quantum effect of the muon zero-point vibrations, which is sizable relative to those of heavier nuclei. The embedded muon, by virtue of its very light mass ($\sim 1/{9^{th}}$ the proton mass), is characterized by zero-point vibration with amplitude typically of the order of 1 Bohr radius~\cite{rath1979}. The neglect of this effect may have two major consequences: inaccurate estimation of the contact hyperfine field,  and/or incorrect identification of muon implantation sites. The former is due to neglect of the space extent of the muon wavefunction, whereas the latter happens when the quantum zero-point vibration energy is comparable with the energy difference between the various implantation sites~\cite{moller2013b,moller2013a,bonfa2015,bonfa2016}.

Earlier approaches towards a quantum-mechanical description of the muon zero-point vibration include calculations within the harmonic approximation~\cite{boxwell1993,moller2013a}. However, the muon potential has been discussed and shown to be anharmonic, for instance by total energy calculations with site exploration algorithms~\cite{rath1979,porter1999,herrero2006,bonfa2015}. Furthermore, a break down of the harmonic approximation takes place when within the range of the muon vibrations the potential is not dominated by the second-order term in its Taylor expansion.

Alternative methods do take into account the anharmonic nature of the crystalline potential. One of them consists in the potential exploration approach~\cite{bonfa2015}. The non-BO methods represent another computationally demanding alternative, employing a linear combination of Gaussian basis functions to realize both the nuclear and the electronic degrees of freedom~\cite{tachikawa1998,hiromi2001,webb2002,Kerridge2004}  and optimized local potentials to represent the nuclear-electron correlation~\cite{gidopoulos2014}. One of the most advanced approaches relies on \emph{ab initio} path integral molecular dynamics, which allows for contextual quantization of both the muon and the electrons in the calculation of the electronic structure and of the interatomic forces ~\cite{valladares1995,probert2006,herrero2006}. However, computational resources required by this method grow exceedingly with the size of the cell. 

In this paper, we describe a stochastic self-consistent harmonic approximation (SSCHA) that allows us to include the effects of anharmonicity in the muon vibrations~\cite{errea2013,errea2014,bianco2017,monacelli2018}. The SSCHA is a quantum variational method that efficiently calculates anharmonic free energies and phonon frequencies in a non-perturbative way. This approach has been very successful for calculating phonon frequencies and superconducting properties in hydrogen-rich materials, as well as in systems undergoing charge density wave (CDW) transitions, ferroelectrics, and thermoelectrics~\cite{errea2013,ion2015a,leroux2015,errea2016,ribeiro2018,unai2019,bianco2019}. For the muon, the SSCHA is variational in the muon (free) energy, with this energy evaluated stochastically from forces and energies calculated at a sufficient number of random muon configurations. The muon energy is minimized using trial harmonic wavefunctions that are Gaussian, while the minimization parameter is the width of the Gaussian. From the output of the minimization, muon frequencies including anharmonic contributions and the muon ground-state energy can be extracted.

With this approach, we demonstrate that there are anharmonic contributions to the harmonic muon vibrational modes, as expected for the muon due to its light mass. We further use the SSCHA muon wavefunction to refine the contact hyperfine field in a series of metals: Fe, Ni, Co, MnSi and MnGe, where the SSCHA improves the agreement of the calculated value with the experimental results, with respect to recent point impurity calculations~\cite{onuorah2018}. Finally, the SSCHA together with energy curvature considerations~\cite{bianco2017} allows the stable occupation of the muon at the octahedral site in bcc Fe, which is unstable within the harmonic regime.

The paper has the following structure: Sec.~\ref{sec:DBO} discusses the double Born-Oppenheimer approximation, which allows us to separate the muon degrees of freedom from those of the host nuclei and electrons. In Sec.~\ref{sec:anharmcalc2}, we describe the working principles of the SSCHA, including the stochastic implementation. In Sec.~\ref{sec:zerofpnt}, we discuss the muon zero-point energy  calculation results using the SSCHA together with the stability of the muon at octahedral and tetrahedral sites in Fe(bcc). Finally, in Sec.~\ref{sec:anharmresults}  we present the results of the quantum corrections in the calculation of the contact hyperfine field and then conclusions are given in Sec.\ref{sec:concl}.

\section{\label{sec:DBO}Double Born-Oppenheimer approximation}

The BO approximation considers the nuclei frozen on the time scale of electron dynamics in view of their sufficiently large mass ratio~\cite{born1927}. Hydrogen is already sufficiently lighter than most other atoms to allow a further separation of time scales, and this holds \emph{a fortiori} true for a positive muon. This allows for the quantum treatment of a single muon \emph{impurity} in the crystal by employing the so-called double Born-Oppenheimer approximation (DBO)~\cite{soudackov1999,porter1999,bonfa2015}. The muon dynamics ($m_\mu \sim 200 m_e$) is much slower than that of electrons, thus justifying an electron structure obtained by DFT with frozen muon and nuclei. The same muon dynamics is still much faster than that of other nuclei, since transition metals are typically 400 times heavier than a muon (care must be taken when considering e.g., hydrogen, which is only nine times heavier than a muon). Therefore it is justified to use total DFT energy versus the muon configuration coordinates as a frozen potential energy landscape in which the muon dynamics takes place on its characteristic time scale. This allows us to consider the zero-point vibration of only the muon within the potential energy surface, \emph{drastically reducing the computational load requirements for the calculations}.

The total Hamiltonian $H_{tot}$ describing the many-body interaction including explicitly the muon coordinates is written as
\begin{equation}
H_{tot} = T_{e} + T_{\mu} + T_{N} + V(\mathbf{r}_e, \mathbf{r}_\mu,  \mathbf{R}_N),
\end{equation}
with subscript $\mu$ describing the muon-related quantities while $e$ and $N$ describe those of the electrons and host nuclei respectively. $T$ and $V$ are the kinetic and potential energy, respectively. The Schr\"odinger equation is then written as
\begin{equation}
H_{tot}  \left| \Psi_{tot} \right>=  E_{tot}  \left| \Psi_{tot}\right>.
\end{equation}
This further allows us to write the DBO wavefunction as a product wavefunction of the electrons, the muon and the nuclei in the form:
\begin{equation}
 \left| \Psi_{tot} \right>=  \left| \psi_{e}\right> \left| \phi_{\mu}\right> \left| \Phi_{N}\right>.
\end{equation}
The Hamiltonian for the electronic problem can be re-written to specifically point out the presence of the muon position operator as
\begin{equation}
H_e = T_{e} + V(\mathbf{r}_e; \mathbf{r}_\mu,  \mathbf{R}_N).
\label{eq:electrohamil}
\end{equation}
Similar to the BO approximation, only the position operators of the muon and the nuclei enter in the eigenvalue problem of the electrons.  The solution of the electronic problem gives the BO  potential energy surface, $V(\mathbf{r}_\mu,  \mathbf{R}_N)$, dependent on the muon and the nuclei position operators. 
 
Hence, the ground-state Hamiltonian $H_\mu$ for the muon can be written as
\begin{equation}
H_\mu = T_{\mu} + V( \mathbf{r}_\mu;  \mathbf{R}_N),
\label{eq:nonanalyham}
\end{equation}
where the muon kinetic energy $T_{\mu} $ is defined as
\begin{equation*}
T_{\mu} = \sum_{\lambda=1}^{3} {\frac{p_\lambda^2}{2m_\mu}} ,
\label{eq:momopera}
\end{equation*}
with $p$ the momentum operator along the Cartesian component indexes $\lambda$, while $m_\mu$ is the muon mass.

The acquisition of the DBO  potential energy surface $V( \mathbf{r}_\mu; \mathbf{R}_N)$  for the solution of the Schr\"odinger equation Eq.~\ref{eq:nonanalyham}, is still a long and difficult task. However, the DBO approximation is advantageous since it allows to consider separately only the degrees of freedom of the muon.
For this reason, in the next section we revisit the SSCHA theory originally presented in Ref.~\cite{errea2013,errea2014} specializing its application to the muon dynamics.

\section{\label{sec:anharmcalc2}Stochastic self-consistent harmonic approximation (SSCHA) for muons}
To begin with the formal description of the SSCHA restricted only to the muon modes, let us write the muon Hamiltonian $H_\mu$, the muon wavefunction $\phi_{\mu}$, and the DBO potential energy surface $V( \mathbf{r}_\mu;  \mathbf{R}_N)$, appearing in the previous section, simply as $H$, $\phi$ and $V(\mathbf{r}_\mu)$ respectively. 

The muon zero-point energy from the Hamiltonian $H$ is given as
\begin{equation}
E= \left< \phi \left| H \right| \phi \right> \label{eq:egrd},
\end{equation}
where $\left| \phi\right>$ is the muon ground-state wavefunction.
Calculating $E$ is far from trivial since the form of the muon potential (Eq.~\ref{eq:nonanalyham}) is not known. However, it is possible to establish a quantum variational principle for the muon ground state energy $E$, by replacing the exact muon wavefunction $\left| \phi\right>$ with the wavefunction  $\left|{\widetilde{\phi}}\right>$  of a trial muon Hamiltonian $\widetilde{H} = T_{\mu} + \widetilde{V}(\mathbf{r}_\mu)$ with energy
\begin{equation}
{\widetilde{E}}= \left< {\widetilde{\phi} } \left| \widetilde{H} \right|{\widetilde{\phi} } \right>. 
\label{eq:trial_harmonic}
\end{equation}
This is such that one can define an energy functional of the trial Hamiltonian as
\begin{equation}
\widetilde{E}_H[\widetilde{H}]= \left<{\widetilde{\phi} } \left| H \right| {\widetilde{\phi} } \right>.
\label{eq:einvar}
\end{equation}
The variational form of the muon ground state energy can be written as
\begin{equation}
E   \leq   \widetilde{E}_H[\widetilde{H}]
\label{eq:variation}
\end{equation}
such that the equality holds when the true and trial potentials are the same.

By adding and subtracting Eq.~\ref{eq:trial_harmonic} to Eq.~\ref{eq:einvar}, $\widetilde{E}_H[\widetilde{H}]$ is written in the form
\begin{equation}
\widetilde{E}_H[\widetilde{H}] = {\widetilde{E}} + \left< {\widetilde{\phi} } \left|(V-\widetilde{V})\right|{\widetilde{\phi} } \right>. 
\label{eq:mue}
\end{equation}
The above definitions allow to formulate a variational principle following the  Gibbs-Bogoliubov inequality theorem~\cite{isihara1968} at zero temperature, similar to the Rayleigh-Ritz inequality~\cite{macdonald1933}. 

According to the trial wavefunction, the probability to find the muon in the position $\mathbf{r}_\mu$ is  
\begin{equation}
{\widetilde{\rho}}(\mathbf{r}_\mu) = \left< \mathbf{r}_\mu \left|  {\widetilde{\phi} } \right> \left<  {\widetilde{\phi}} \right| \mathbf{r}_\mu \right> =  \left| {\widetilde{\phi}}(\mathbf{r}_\mu)\right|^2.
\label{eq:prbdenss}
\end{equation}
Thus, an observable \textit{A} dependent only on $\mathbf{r}_\mu$ can be averaged statistically within the form of the corresponding Hamiltonian $\widetilde{H}$ as
\begin{equation}
\left<\mathit{A}\right>_{\widetilde{H}}= \int{d\mathbf{r}_\mu \mathit{A}(\mathbf{r}_\mu){\widetilde{\rho}}(\mathbf{r}_\mu)}
\label{eq:statav}
\end{equation}
and the muon energy  in Eq.~\ref{eq:mue} can be evaluated as  
\begin{equation}
\widetilde{E}_H[\widetilde{H}] = {\widetilde{E}} + \int{d\mathbf{r}_\mu {\widetilde{\rho}} (\mathbf{r}_\mu) (V(\mathbf{r}_\mu)-\widetilde{V}(\mathbf{r}_\mu))}.
\label{eq:enegcal}
\end{equation}
With the above form of $\widetilde{E}_H[\widetilde{H}] $, the muon energy can be evaluated at each step during the variational minimization. One can directly see that the equality in the form of the variation in Eq.~\ref{eq:variation} holds if $V = \widetilde{V}$.  Hence, with the variational principle,  the ground state of the muon is determined if the potential $\widetilde{V}(\mathbf{r}_\mu)$ that minimizes $\widetilde{E}_H[\widetilde{H}]$ is found. 

To proceed with the minimization of $\widetilde{E}_H$, in the SSCHA implementation we restrict the muon wavefunctions only to the Gaussian form. The term \emph{harmonic} in the technique refers to the fact that each Gaussian is the ground state of a trial \emph{harmonic} Hamiltonian, with known analytic solutions (see Appendix~\ref{sec:harmonicharm}) where the trial potential is expressed in terms of a force constant matrix. 
Moreover, using Gaussian functions has the advantage of allowing to sample the wavefunction by extracting randomly distributed configurations without any Metropolis algorithm that requires long equilibration time and also provides an analytic expression for the kinetic energy.

\begin{table*}[!hbtp]
\centering
\caption[Muon effective harmonic frequencies and zero-point energy]{\label{tab:efffzp} Harmonic muon frequencies $\omega_i^h$ along the mode $i$ and harmonic zero-point energy $E^h=\sum_{i=1}^{3} \hbar \omega_{i}^h/2$, together with the SSCHA muon frequencies $\widetilde{\omega}_{i}$ and energy $\widetilde{E}$ at the minimum that includes the anharmonic contribution.} 
\begin{ruledtabular}
 \begin{threeparttable}
 \begin{tabular}{ l c c c c c c c c c c}
    Host   & $\omega^h_x$ (cm$^{-1}$)  & $\omega_y^h$ (cm$^{-1}$)& $\omega^h_z$ (cm$^{-1}$) & $E^h$ (eV) & $\widetilde{\omega}_x $ (cm$^{-1}$)  & $\widetilde{\omega}_y$ (cm$^{-1}$)& $\widetilde{\omega}_z$ (cm$^{-1}$) & $ \widetilde{E} $ (eV) \\ 
      \hline
    Fe - bcc~\tnotex{tnxx6} & 4364.01& 2913.01 &  4364.62  &  0.72   & 4769.08 & 2572.58 &  5088.37      &0.74 \\ 
    Fe - bcc~\tnotex{tnxx62}  & 1965.08$i$& 1958.72$i$& 6828.00    &  ~\tnotex{tnxx63} & 2005.24  & 2005.24 &  6364.81      &0.53 \\ 
    Co - hcp & 2930.41& 2929.85& 2752.25   & 0.53     & 3741.10 & 3741.10  &3476.24       &0.61 \\
    Co - fcc  &2607.29& 2607.02& 2606.66   & 0.49 &3424.16 & 3424.16  &3424.16      &0.56 \\
    Ni - fcc   & 2377.62& 2377.60& 2377.61    &  0.44   &3317.78  &3317.78   & 3317.78      &0.53\\ 
    MnGe    & 3123.70& 3123.67& 3123.66  &0.58 &3470.29  & 3470.29  &3470.29      &0.64 \\
    MnSi   & 3296.27& 3296.32&  3296.11  & 0.61   &3685.25   &3685.25   &3685.25    &0.67 \\ 
    \end{tabular}
   \begin{tablenotes}
            \item[a]  \label{tnxx6} Muon at the tetrahedral site.
             \item[b]  \label{tnxx62} Muon at the octahedral site.
             \item[c]  \label{tnxx63} The muon is not stable at the octahedral site (imaginary frequencies) within the harmonic regime.
 \end{tablenotes} 
  \end{threeparttable}
\end{ruledtabular}
\end{table*}

Finally, the actual minimization is obtained using the conjugate gradient (CG) algorithm~\cite{hestenes1952} which requires the evaluation of the energy gradient, whose analytic form is given in Ref.~\cite{errea2014} and in Appendix~\ref{sec:enestoch2} for the muon case, and depends on the forces acting on the muon when displaced from the equilibrium position.

The evaluation of the quantities of interest at each minimization step, namely $\widetilde{E}_H$ and its gradient, is performed stochastically. One of the advantages of the stochastic sampling resides in the gradual optimization of the potential felt by the muon during the iterative process. This ensures that the entire BO landscape, beyond the harmonic component around the minimum, is sampled, hence capturing the anharmonic effects.  

The stochastic sampling of the BO energy and of the forces acting on the muon and entering the energy gradient (see Appendix~\ref{sec:enestoch2}) can be calculated with any \emph{ab initio} method including DFT~\cite{kohn1965} and Hartree-Fock~\cite{fock1930,hartree1947,slater1951} based approaches. 

The evaluation of the forces and energies for the random muon configurations in the stochastic sampling represents the most computationally demanding task in the SSCHA minimization cycle. This effort can be partially alleviated with a re-weighting procedure based on importance sampling. The reader is referred to Ref~\onlinecite{errea2014} for a detailed description of this additional detail.

When the energy gradient numerically vanishes, the $\widetilde{E}$ that minimizes $\widetilde{E}_H[\widetilde{H}]$ is the zero-point energy of the muon and the anharmonic vibrational frequencies $\widetilde{\omega}_{i}$ 
of the auxiliary Hamiltonian whose SSCHA wavefunction is the ground state are obtained, so that
\begin{equation}
{\sum_{i=1}^{3}{ \frac{1}{2}\hbar \widetilde{\omega}_{i}=\widetilde{E}}}.
\end{equation}

The formal description of the trial Hamiltonian and the trial wavefunction is given in Appendix A.

  
\begin{figure}[!h]
\centering
\textbf{(a)} \\
\centering
\includegraphics[width=7.8cm]{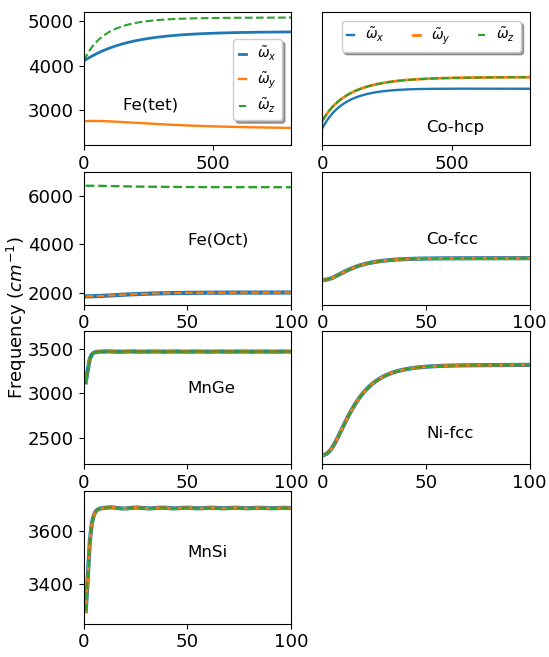} \\
\centering
\textbf{(b)} \\
\centering
\includegraphics[width=7.8cm]{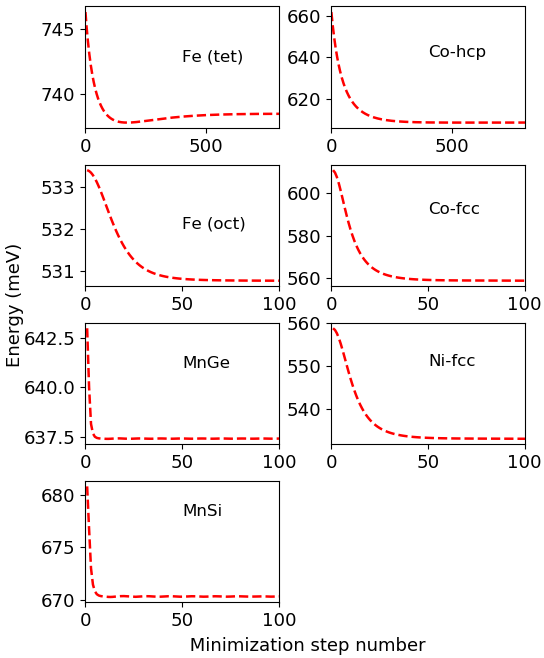} \\
\caption[Frequency and Energy evolution]{\label{fig:freQ Energy Evolution} (a) Evolution of the SSCHA muon frequency during the minimization steps  (b) Evolution of SSCHA muon energy as in Eq.~\ref{eq:enegcal} during the minimization steps. In both figures the starting point for the minimization step number = 0, is that of the harmonic Hamiltonian except for the muon in octahedral site of Fe (Fe (oct)) when the starting potential is arbitrary.}
\end{figure} 
\section{\label{sec:zerofpnt}Muon zero-point energy}
 Let us first describe the zero-point energy of the muon obtained in the harmonic approximation, which is later used in comparisons with the anharmonic one.

The harmonic muon frequencies  $\omega^h_i$ and the corresponding energies $E^h=\sum_{i=1}^{3} \hbar \omega_{i}^h/2$ were calculated by the finite difference method~\cite{kresse1995,parlinski1997}, which allows only the muon frequencies to be singled out, for all the materials under investigation, namely Fe, Co, Ni, MnSi and MnGe. 
These were also used to generate the starting wavefunctions for the SSCHA minimization except for the stability discussion in Sec.~\ref{sec:tetoct} with the muon at the octahedral and tetrahedral site in bcc Fe. Here, the density functional perturbation theory (DFPT) within the Quantum ESPRESSO suite of code~\cite{baroni2001,qe2009} was used to calculate the frequencies of the whole system, including those of the host Fe nuclei. The resulting harmonic muon frequencies from both methods in the two Fe systems are in good agreement.

For the SSCHA minimization and stochastic averaging (see Eq~\ref{eq:statavstoch}), hundreds (100 to 400) of random configurations were generated for the muon, while keeping the host atoms fixed, to ensure that the muon energy gradient vanishes.  Their energies and  Hellmann-Feynman forces~\cite{pulay1969} were calculated by DFT as implemented in the Quantum ESPRESSO suite of code~\cite{qe2009}. The details of the muon site in these systems and DFT input parameters are contained in Ref.~\cite{onuorah2018}. For all the systems, a 2$\times$2$\times$2 supercell constructed starting from the conventional unit cell was used for the harmonic frequency calculations, the SSCHA frequency minimization and the force calculation within DFT. Other DFT computational details are identical to those reported in Ref.~\cite{onuorah2018}. To accommodate the muon impurity in the supercell, the forces introduced by the muon in the system were relaxed by DFT and the relaxed structures were used for the SSCHA calculations. Relaxations were converged with force and energy thresholds of 10$^{-3}$ a.u and 10$^{-4}$ Ry respectively.

Figure \ref{fig:freQ Energy Evolution} shows the evolution of the SSCHA muon frequencies and energy during the minimization procedure. Significant anharmonic contributions to the resulting SSCHA frequencies can be deduced from the difference between the initial values, i.e. the starting harmonic guess, and the final converged results (the comparison with the anharmonic correction obtained for host atoms is presented in Appendix ~\ref{sec:appendixC}). The anharmonic correction to the harmonic frequencies is found to be in the range of 330 - 820 cm$^{-1}$ except for the muon at the octahedral site in Fe. 

The stochastic implementation ensures that the effect of the muon vibrations, the effect of the chemical environment around the muon and anharmonic contributions to the forces acting on the muon are all incorporated in the muon ground state minimum. 

Table~\ref{tab:efffzp} contains the harmonic frequencies $\omega^h_i$ and energies $E^h$, obtained with the finite difference method and used as the starting point of the SSCHA iterative process,  and the SSCHA frequencies  $\widetilde{\omega}_i$ and energies  $\widetilde{E}$  at the end of the minimization. The error estimates of the reported muon energies are within the range of 0.1\,meV. The results show the anharmonic effects in the muon vibrational frequencies. Notice that the muon at octahedral implantation site in Fe is  unstable in the harmonic regime. For all other cases with positive harmonic frequencies for which $E^h$ can be defined, the difference between the SSCHA muon vibrational energies and the harmonic ones is in the range of 0.02 - 0.09\,eV.

\section{\label{sec:tetoct}Tetrahedral and octahedral muon site in Fe}
Conflicting experimental and theoretical studies report the  muon site in Fe to be either at the tetrahedral (T)  or the octahedral (O) interstitial sites~\cite{seeger1976,graf1980,lindgren1982,estreicher1982,yagi1984,kossler1985}. From the point of view of the DFT total  energy, the T site is 0.184\,eV lower that the O site. This would indicate that the T site is the stable one. However, since the calculated muon zero-point energies (above 0.5\,eV) are large relative to the DFT energy difference, the possible population of both sites cannot be excluded.

DFPT calculations of the muon frequencies provide further insight into the stability of the two candidate sites. Unphysical \emph{negative} frequencies, generally a signal of instability, are obtained for the muon at the O site, as opposed to those of the T site, which are always positive. The harmonic approximation then appears to indicate an instability of the muon at the O site. 

However, the anharmonic effects, fully captured by the SSCHA, yield positive frequencies also for the O site indicating that the instability is an artifact of the harmonic approximation. As the $\widetilde{\omega}_i$ frequencies are positive-definite by definition, this is not proof that the O site occupation is stable. Obtaining the frequencies from the energy curvature~\cite{bianco2017}, which can correctly describe an instability, confirms, however, that the O site interstitial site is in fact stable. The SSCHA frequencies for the muon in the O site are larger than the frequencies resulting from those obtained from the curvature by only 0.53 \% along the $x$,$y$ axis and 0.14 \% along the $z$ axis.
 
 The quantum correction with the SSCHA shows that both T and O are stable local minima. The vibrational contribution to the energy is 0.21 eV less for the O site than for the T site (see Table~\ref{tab:efffzp}). Adding this to the static DFT contribution makes the O site energetically favored by approximately 0.03 eV over the T site, thus indicating that the two sites are basically degenerate, and possibly both occupied.

\section{\label{sec:anharmresults}Quantum corrections on a muon contact hyperfine field}

The contact hyperfine field $B_c(\mathbf{r}_\mu)$ at the muon position $\mathbf{r}_\mu$ is computed \emph{ab inito} by considering the imbalance in the spin density at the muon site~\cite{onuorah2018} given as 
\begin{equation}
B_c(\mathbf{r}_\mu) = \frac{2}{3}\mu_0 \mu_B \left[n_\uparrow(\mathbf{r}_\mu)-n_\downarrow(\mathbf{r}_\mu)\right],    
\end{equation}
where $\mu_0$ is the vacuum permeability, $\mu_B$ is the Bohr magneton and $n_\uparrow-n_\downarrow$ represents the spin polarization at the muon site $\mathbf{r}_\mu$ calculated here by DFT. $B_c(\mathbf{r}^{eq}_\mu)$  has been calculated in this way for metals within a point impurity treatment of the muon~\cite{onuorah2018}. We now calculate the effect of the muon quantum delocalization on its contact hyperfine field, using the muon SSCHA wavefunctions $\phi$ that already contain the anharmonic contributions.

The quantum expectation value, $\left<B_c\right>$  is given by
\begin{equation}
\left<B_c\right> = \int{ d\mathbf{r}_\mu B_c(\mathbf{r}_\mu) \left|\phi(\mathbf{r}_\mu)\right|^2}.
\label{eq:muoav1}
\end{equation}

where the probability density $\left|\phi(\mathbf{r}_\mu)\right|^2$ has been defined in Eqs.~\ref{eq:prbdenss} and is obtained from the SSCHA muon frequencies $\widetilde{\omega}_{i}$ according to Eq.  ~\ref{eq:rhoharm}. 

\begin{figure}[!h]
\includegraphics[width=8.0cm]{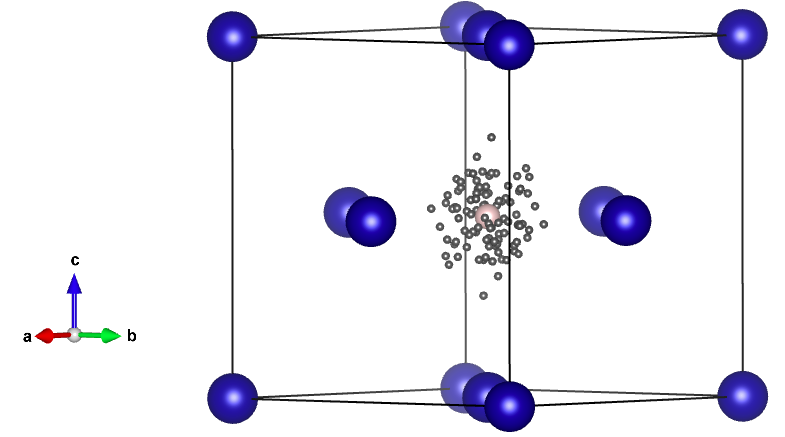}
\centering
\caption[Random points for $B_c$ averaging ]{\label{fig:averagegrid} 100 random position generated using Eq.~\ref{eq:randommupos} for the muon at octahedral site in Co-fcc unit cell. The equilibrium octahedral center is depicted by the pink sphere, while the small dark spheres represent the different random muon positions where the muon contact hyperfine field within point impurity treatment $B_c(\mathbf{r}_\mu)$  was also calculated for the purpose of including the quantum effects of the muon.} 
\end{figure}

The above integral can be evaluated in a post-DFT calculation by a statistical average performed stochastically, i.e. according to

\begin{equation}
\int{d\mathbf{r}_\mu \mathit{B_c}(\mathbf{r}_\mu){{\widetilde{\rho}}}(\mathbf{r}_\mu)} \simeq \frac{1}{N_c} \sum_{n=1}^{N_c} {\mathit{B_c}(\mathbf{r}_{\mu}^{n})} \equiv \left<\mathit{B_c}\right>_{{\widetilde{H}}}.
\label{eq:statavstoch}
\end{equation}
where the sum extends over a number of muon random configurations $N_c$ displaced from the equilibrium position $\mathbf{r}^{eq}_\mu$ and generated with the probability distribution of the muon wavefunction (see Eq.~\ref{eq:randommupos}). The number of muon random configurations used is the same as in the SSCHA minimization of the muon wavefunction. However, the new muon random positions are generated considering the anharmonic corrected SSCHA muon wavefunction. Figure \ref{fig:averagegrid} shows the distribution of the 100  configurations used for fcc Co in the unit cell.

$B_c(\mathbf{r}_\mu)$ was calculated by DFT for each of these random configurations within a 3$\times$3$\times$3 supercell for Fe, Co and Ni and 2$\times$2$\times$2 supercell for MnGe and MnSi, while other computational details are the same as reported in Ref.~\cite{onuorah2018}.

\begin{figure}[!htb]
\includegraphics[width=8.0cm]{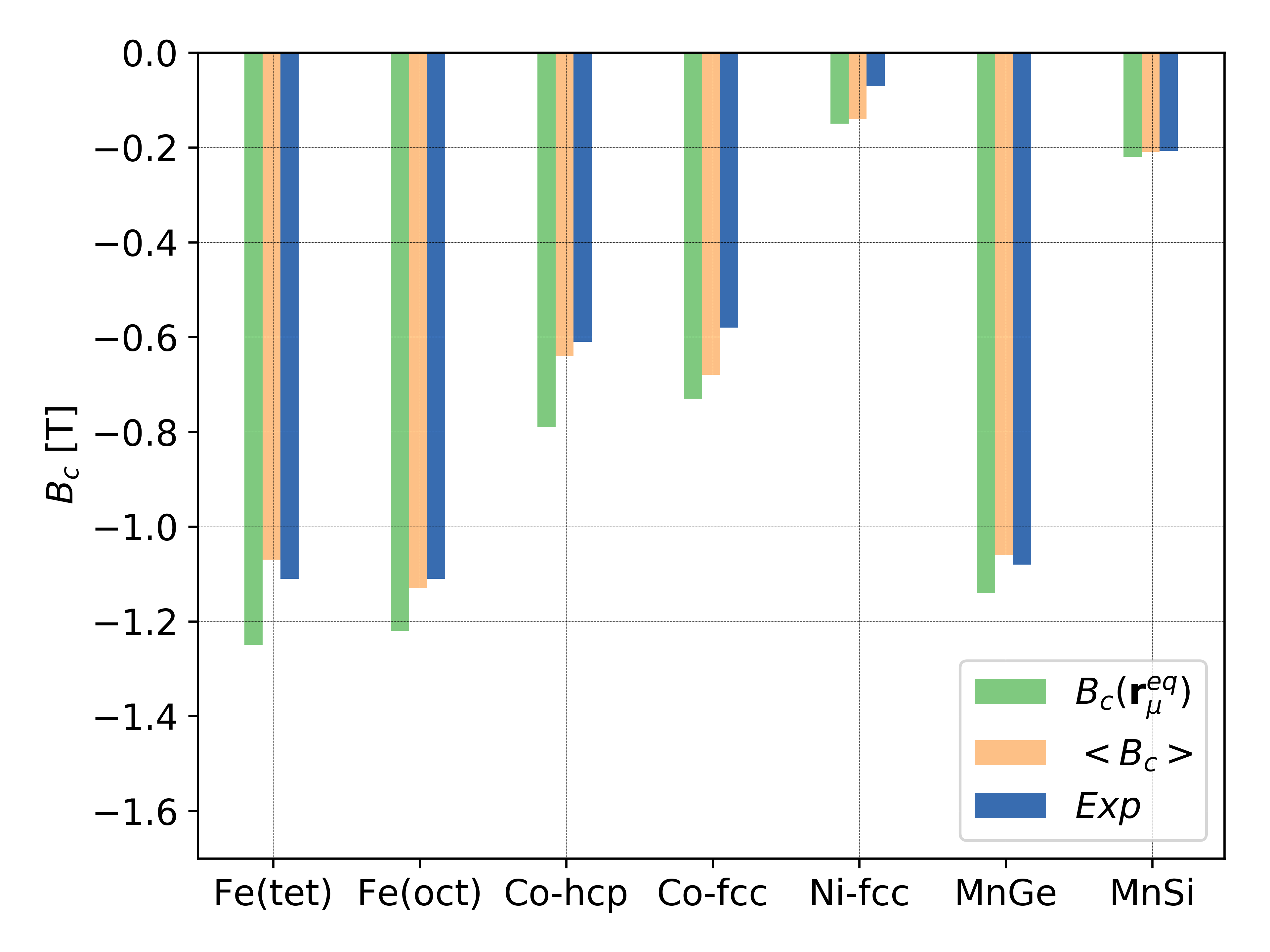}
\centering
\caption[Averaged muon contact hyperfine field]{\label{fig:cont1} Contact hyperfine field $B_c(\mathbf{r}_\mu^{eq})$ at the equilibrium muon implantation position $\mathbf{r}_\mu^{eq}$, the  muon contact field averaged  over the muon wavefunction spread, $\left<B_c\right>$ and experimentally observed values. }
\end{figure}

Table~\ref{tab:averaged_bc} and Fig.~\ref{fig:cont1} show the calculated contact field $B_c(\mathbf{r}^{eq}_\mu)$ for a point-like muon~\cite{onuorah2018} and its stochastically \emph{averaged} $\left<B_c\right>$ values together with the experimental values. For all the systems the statistical error for the stochastic sampling of  $\left<B_c\right>$ is in 
the range of $\approx$~1\,mT. The contact hyperfine field including quantum correction within the SSCHA,  $\left<B_c\right>$, improves the agreement with the experiments, thus underlining the importance of considering the finite muon wavefunction when computing muon hyperfine interactions. Admittedly, the correction to the contact hyperfine field appears to be less relevant than the outcome obtained on the stability of the muon at the octahedral site in Fe, still $|\left<B_c\right>|$ introduces a correction that ranges between 1 and 18\%.

\begin{table} [!htb]
\centering
\caption[Calculated $\mathbf{B}_c$ averaged over the muon wavefunction spreading]{\label{tab:averaged_bc}  Calculated contact hyperfine field for the point muon at the equilibrium position $B_c(\mathbf{r}_\mu^{eq})$, the calculated contact hyperfine field averaged over the spread of the muon wavefunction $\left<B_c\right>$ and experimentally observed values (Exp).}
\begin{ruledtabular}
 \begin{threeparttable}
 \begin{tabular}{ c c c c}
   Host metals & $B_c(\mathbf{r}_\mu^{eq})$ [T]~\tnotex{tnxx667}  & $\left<B_c\right>$ [T] & Exp  \\ 
   \hline
    Fe-bcc~\tnotex{tnxx666}  &-1.25   & -1.07 & -1.11~\cite{nishida1977} \\
     Fe-bcc~\tnotex{tnxx668}  &-1.22   & -1.13 & -1.11~\cite{nishida1977} \\
    Co-hcp  &-0.79  & -0.64 & -0.61~\cite{graf1976} \\ 
    Co-fcc   &-0.73  & -0.68& -0.58~\cite{lindgren1982} \\ 
    Ni-fcc   & -0.15 & -0.14 & -0.071~\cite{graf1979} \\ 
    MnGe & -1.14 & -1.07 & -1.08~\cite{martin2016} \\
    MnSi & -0.22 & -0.21 &  -0.207~\cite{amato2014} \\
\end{tabular}
\begin{tablenotes}
     \item[a]  \label{tnxx667}Ref.~\cite{onuorah2018}
    \item[b]  \label{tnxx666}Muon at the tetrahedral site
    \item[c]  \label{tnxx668}Muon at the octahedral site
 \end{tablenotes}  
 \end{threeparttable}
 \end{ruledtabular}
\end{table}

\section{\label{sec:concl}Conclusion}

In conclusion, we have presented a general, effective and robust approach, based on the DBO approximation, to obtain the ground state wavefunction and zero-point energy of a positive muon embedded in a crystal from first principles. 
The adaptation of the SSCHA to the muon case allows us to evaluate the delocalized muon wavefunction including anharmonic contributions, that correct harmonic ones.

Moreover, the SSCHA circumvents the problem of directly reconstructing the  potential energy surface by replacing this task with a variational problem, and more importantly, it provides a computationally tractable method to describe the zero-point energy of the muon. This leads to a number of important insights concerning the stability of the muon sites and its coupling with the surrounding electrons.

The first point has been discussed by considering the case of the muon site in Fe, where anharmonicity plays a crucial role in establishing the stability of the muon in the tetrahedral and octahedral sites.

We reformulated the calculation of the muon contact hyperfine field by including the effects of its anharmonic zero-point vibration, improving the agreement with experiments with respect to previous estimates based on the point impurity treatment of the muon. Even though the correction is small, in numerous cases the contact field is of the order of tenths of a Tesla, thus making the absolute value of the correction presented here quite relevant.

Finally, the clean iterative procedure of the SSCHA makes it rather straightforward to define standardized workflows to automate the computational procedure. This represents another step towards routinely supporting experimental data analysis with computational simulation results.

\begin{acknowledgments}
RDR acknowledges grants from the European Union's Horizon 2020 research and innovation program under Grant Agreement No. 654000. RDR, PB and IJO also acknowledge computing resources provided by, the Swiss National Supercomputing Centre (CSCS) under Project ID sm16, CINECA under  Project ID IsC58, the STFC Scientific Computing Department’s SCARF cluster and the HPC resources at the University of Parma, Italy. IE acknowledges funding from the Spanish Ministry of Economy and Competitiveness (FIS2016-76617-P). This work is part of the PhD thesis of IJO at the University of Parma, Italy.

\end{acknowledgments}
\appendix
 \section{\label{sec:harmonicharm} The trial muon harmonic Hamiltonian}
The trial muon harmonic Hamiltonian is of the form
 \begin{equation}
 \widetilde{H} = \sum_{\lambda=1}^{3} {\frac{p_\lambda^2}{2m_\mu}} + \frac{1}{2} \sum_{\lambda \nu}^{3}{K^{\lambda \nu}({r}_\mu - {r}_\mu^{eq})^{\lambda} ({r}_\mu - {r}_\mu^{eq})^{\nu}},
 \label{eq:trialharmpot}
 \end{equation}
where $\lambda$ and $\nu$ are Cartesian component indexes, $\mathbf{r}_\mu^{eq}$ is the muon equilibrium position, $m_\mu$ is the mass of the muon and $K^{\lambda \nu}$ is the muon $3 \times 3$ force constant matrix. The force constant matrix  $K^{\lambda \nu}/ m_\mu$ can be constructed and diagonalized as
 \begin{equation}
 \sum_{\nu=1}^{3}{\frac {K^{\lambda \nu}}{ m_\mu} \widetilde{\epsilon}_{i}^{\nu} =  \widetilde{\omega}_{i}^2 \widetilde{\epsilon}_{i}^{\lambda} }, 
\label{eq:frcnst} 
 \end{equation}
where $i$ is the index  of each of the orthogonal modes, $\widetilde{\epsilon}_{i}^{\nu}$ is the polarization vector and $\widetilde{\omega}_{i}$ is the muon frequency corresponding to the trial Hamiltonian $\widetilde{H}$ for each mode.  

Assuming a trial harmonic potential, the probability to find the muon at $\mathbf{r}_\mu$ can be written simply as
 \begin{equation}
 \widetilde{\rho}(\mathbf{r}_\mu) =   \frac{1}{ \prod\limits_{i=1}^{3}  \sqrt{2\pi \widetilde{\sigma}_{i}^{2} } }  \exp \left(- \sum_{\lambda \nu i}^{3}{ \frac{\widetilde{\epsilon}_{i}^{\lambda} \widetilde{\epsilon}_{i}^{\nu}}{2\widetilde{\sigma}_{i}^{2} }({r}_\mu - {r}_\mu^{eq})^{\lambda} ({r}_\mu - {r}_\mu^{eq})^{\nu}} \right),
\label{eq:rhoharm}
\end{equation}
where $\widetilde{\sigma}_{i}$, the normal length for  each of the modes $i$, is given as
\begin{equation}
\widetilde{\sigma}_{i} = \sqrt{\frac{\hbar}{2m_\mu \widetilde{\omega}_{i}}}.
\label{eq:sigma}
\end{equation}
Using the quantum statistical averaging defined in Eq.~\ref{eq:statav}, the energy of the trial harmonic Hamiltonian can be calculated as
\begin{equation}
{\widetilde{E}} = \sum_{i=1}^{3}{ \frac{1}{2}\hbar \widetilde{\omega}_{i}}.
\end{equation}

\section{\label{sec:enestoch2} Random configuration sampling and  minimization details}
The distribution for the generation of the random muon position configurations is realized using  random numbers $ \{\xi_{in}\}_{n=1,...,{N_c}} $  generated with the Gaussian distribution ${\widetilde{\rho}}(\mathbf{r}_\mu)$  and re-scaled by the corresponding normal length modes $\widetilde{\sigma}_{i}$ and polarization vector $\widetilde{\epsilon}_{i}^{\lambda}$. The generated positions are thus obtained as
\begin{equation}
(r_{\mu}^{n})^{\lambda} = (r_{\mu}^{eq})^{\lambda} + \sum_{i=1}^{3}{\widetilde{\epsilon}_{i}^{\lambda}\widetilde{\sigma}_{i}\xi_{in}}.
\label{eq:randommupos}
\end{equation}

This constitutes the set of points used in the stochastic evaluation of $\widetilde{E}$ and of the gradient of the energy functional, namely $\nabla_{K}\widetilde{E}_H[\widetilde{H}]$, with respect to the force constant $K$. The analytic form of this last term is written as (see also Ref.~\cite{errea2014})
\begin{equation}
\begin{split}
\nabla_{K}\widetilde{E}_H[\widetilde{H}] =& - \sum_{i \lambda \nu}{\left(\widetilde{\epsilon}_{i}^{\lambda}\nabla_K \ln \widetilde{\sigma}_{i}   +  \nabla_K \widetilde{\epsilon}_{i}^{\lambda}\right) \widetilde{\epsilon}_{i}^{\nu}}   \times \\
& \int { d \mathbf{r}_\mu [ \mathit{f}^{\lambda} (\mathbf{r}_\mu) - {\widetilde{\mathit{f}}}^{\lambda} (\mathbf{r}_\mu)] (r_\mu - r_\mu^{eq})^\nu {\widetilde{\rho}}(\mathbf{r}_\mu)},
\end{split}
\label{eq:grad2}
\end{equation} 
where $\mathit{f}^{\lambda} (\mathbf{r}_\mu) $ is the muon force component in the $\lambda$ Cartesian direction  for all muon positions  $\mathbf{r}_\mu$ and ${\widetilde{\mathit{f}}}^{\lambda} (\mathbf{r}_\mu)$ are the forces obtained with the ${\widetilde{V}}$ potential. The SSCHA minimization is performed respecting the symmetries of the crystal~\cite{errea2014}.

We also add that with the SSCHA, it is possible to minimize the energy both with respect to the muon position $\mathbf{r}_\mu$ and also the force constant matrix $K^{\lambda \nu}$. However, for the materials considered in this paper, there is sufficient knowledge of the equilibrium muon position $\mathbf{r}_\mu^{eq}$. Hence, the muon energy is only minimized with respect to the force-constant matrix $K$. For the muon in a high symmetry position, the force-constant matrix is a 3$\times$3 matrix, with the diagonal elements of the matrix accounting for the dominant contribution. 

Finally, it is important to note that, in order to obtain physical phonons from the ground-state minimized quantities provided by the SSCHA, the second derivative (curvature) of SSCHA energy at the minimum with respect to $\mathbf{r}_\mu$ has to be calculated~\cite{bianco2017}, which includes a correction term to the force constants matrix $K^{\lambda \nu}/ m_\mu$ in Eq.~\ref{eq:frcnst}. We verified that for the cases under study here the muon frequencies are affected by less than a 1\% by this extra correction. Thus, we can  treat the $\widetilde{\omega}_{i}$ frequencies as the physical vibrational energies of the muons.

\section{\label{sec:appendixC} Evolution of muon and host Fe SSCHA frequencies}
\begin{figure}[h!]
\includegraphics[width=7.5cm]{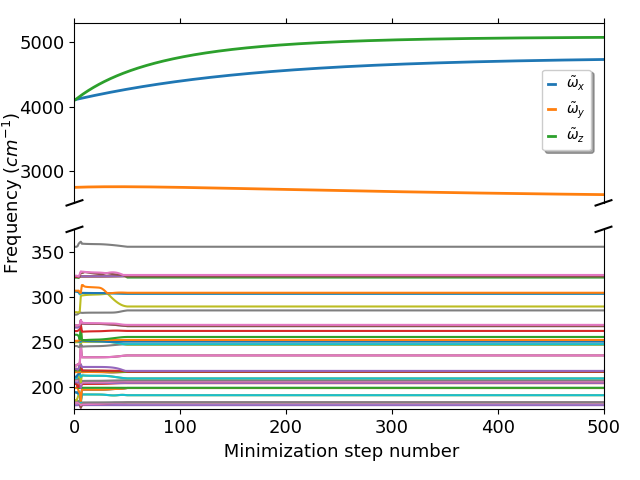}
\centering
\caption[Evolution of muon and Fe frequencies]{\label{fig:evolu1} Evolution of the SSCHA muon frequency ($\tilde{\omega}$ in the upper panel) and those of Fe (nearly static low frequency lines in the lower panel) during minimization for muon in tetrahedral site of bcc Fe. The figure depicts the expected anharmonicity effects on the SSCHA muon frequencies and nearly non-existent anharmonicity effects on those of Fe, due to the large mass difference of the muon and Fe nuclei. The muon is $\approx$ 490 times lighter.}
\end{figure}

The evolution of the frequencies in the SSCHA calculation including anharmonic effects both for the Fe host nuclei and the muon at the tetrahedral site in a 2$\times$2$\times$2 supercell is shown in Fig. \ref{fig:evolu1}. The figure indicates that there is a significant anharmonic contribution to the muon eigenfrequencies after several iterations (upper panel), whereas the lower frequency modes of the heavier Fe nuclei (lower panel) remain negligibly changed. This consideration together with  the DBO approximation discussed in sec.~\ref{sec:DBO} also supports separating and concentrating only on the muon degrees of freedom.

\bibliography{references}

\end{document}